\documentclass[a4paper,11pt]{article}
\pdfoutput=1 % if your are submitting a pdflatex (i.e. if you have
\usepackage{jinstpub} % for details on the use of the package, please
\usepackage{comment}
\usepackage{graphicx}
\usepackage{notoccite}
\usepackage{hyperref}

\usepackage{caption}
\usepackage{subcaption}

\newcommand{\addComment}[2]{
  \expandafter\newcommand\csname #1\endcsname[1]{{\bf \color{#2} \capitalisewords{#1}:\,##1}}
  \expandafter\newcommand\csname #1cor\endcsname[2]{{\color{#2} \capitalisewords{#1}:\,\st{##1}{\bf ##2}}}
  \expandafter\newcommand\csname #1color\endcsname{#2}
}

\newcommand{\GlueX}{\textsc{GlueX}}

\addComment{cris}{red} 
\addComment{azizah}{magenta} 
\title{Artificial Intelligence for Imaging Cherenkov Detectors at the EIC}

\author[a,*]{C. Fanelli}
\author[c]{A. Mahmood}

\affiliation[a]{ Laboratory for Nuclear Science, Massachusetts Institute of Technology, Cambridge, MA 02139, U.S.A.}
\affiliation[b]{University of Regina, Regina, SK S4S 0A2, Canada}

\emailAdd{cfanelli@mit.edu}

\abstract{Imaging Cherenkov detectors form the backbone of particle identification (PID) at the future Electron Ion Collider (EIC). 
 Currently all the designs for the first EIC detector proposal use a dual Ring Imaging CHerenkov (dRICH) detector in the hadron endcap, a Detector for Internally Reflected Cherenkov (DIRC) light in the barrel, and a modular RICH (mRICH) in the electron endcap.  
 These detectors involve optical processes with many photons that need to be tracked through complex surfaces at the simulation level, while for reconstruction they rely on pattern recognition of ring images. 
 This proceeding summarizes ongoing efforts and possible applications of AI for imaging Cherenkov detectors at EIC. 
 In particular we will provide the example of the dRICH for the AI-assisted design and of the DIRC for simulation and particle identification from complex patterns and discuss possible advantages of using AI. 
 }


\arxivnumber{} % only if you have one

\proceeding{1$^{\text{st}}$ Workshop on Artificial Intelligence for the Electron Ion Collider\\
  September 7-10, 2021\\
  Center for Frontiers in Nuclear Science}

\begin{document}
\maketitle
\flushbottom

\section{Introduction}
\label{sec:intro}

 In the proposed Electron-Ion Collider (EIC), imaging Cherenkov detectors are the backbone of particle identification. When charged particles move through the dielectric media of the Cherenkov detector at a speed larger than the phase velocity of light in that medium, they emit Cherenkov radiation in a characteristic conical shape. 
 
\begin{figure}[!htbp]
\centering 
\includegraphics[width=.40\textwidth]{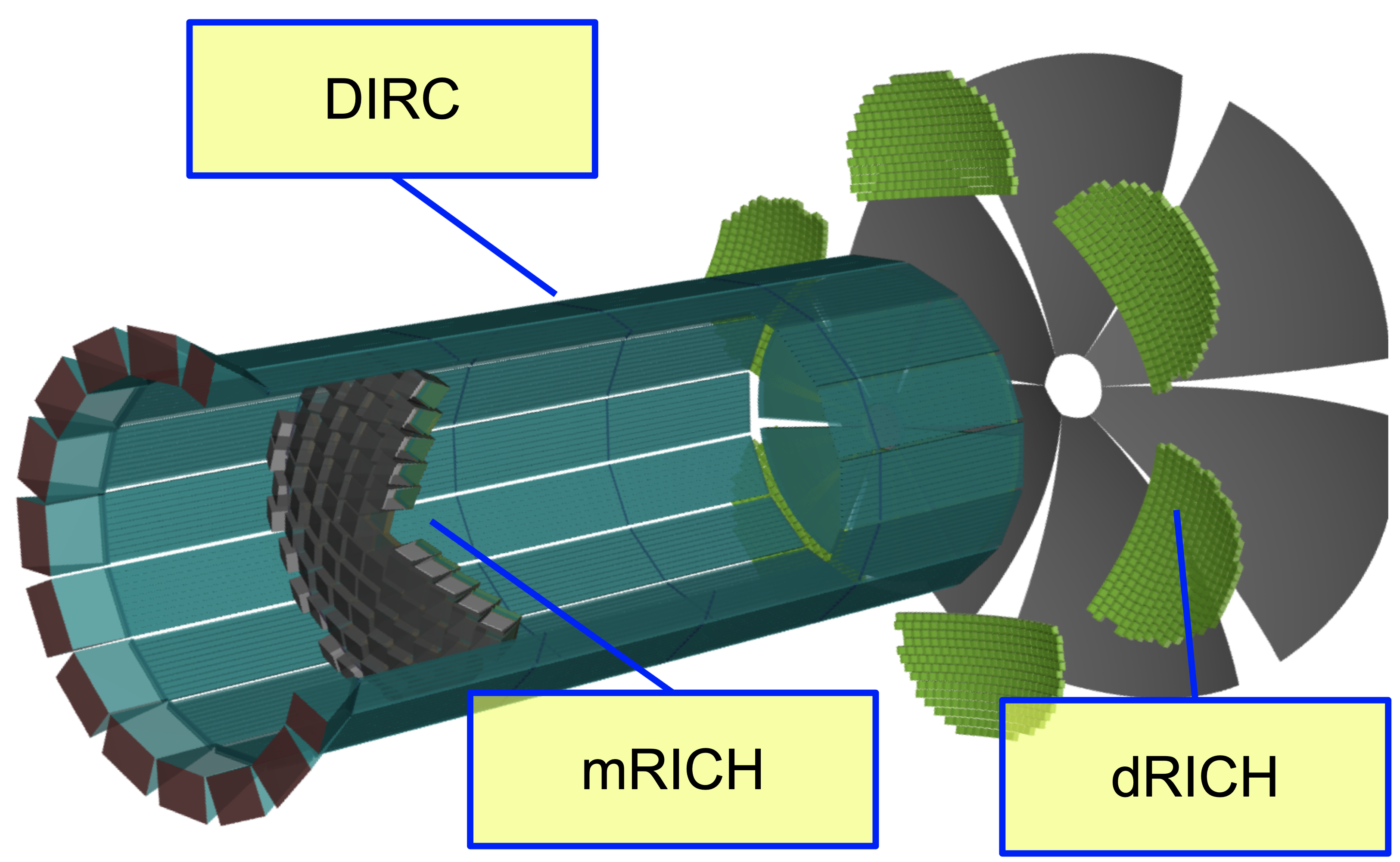}
\includegraphics[width=.50\textwidth]{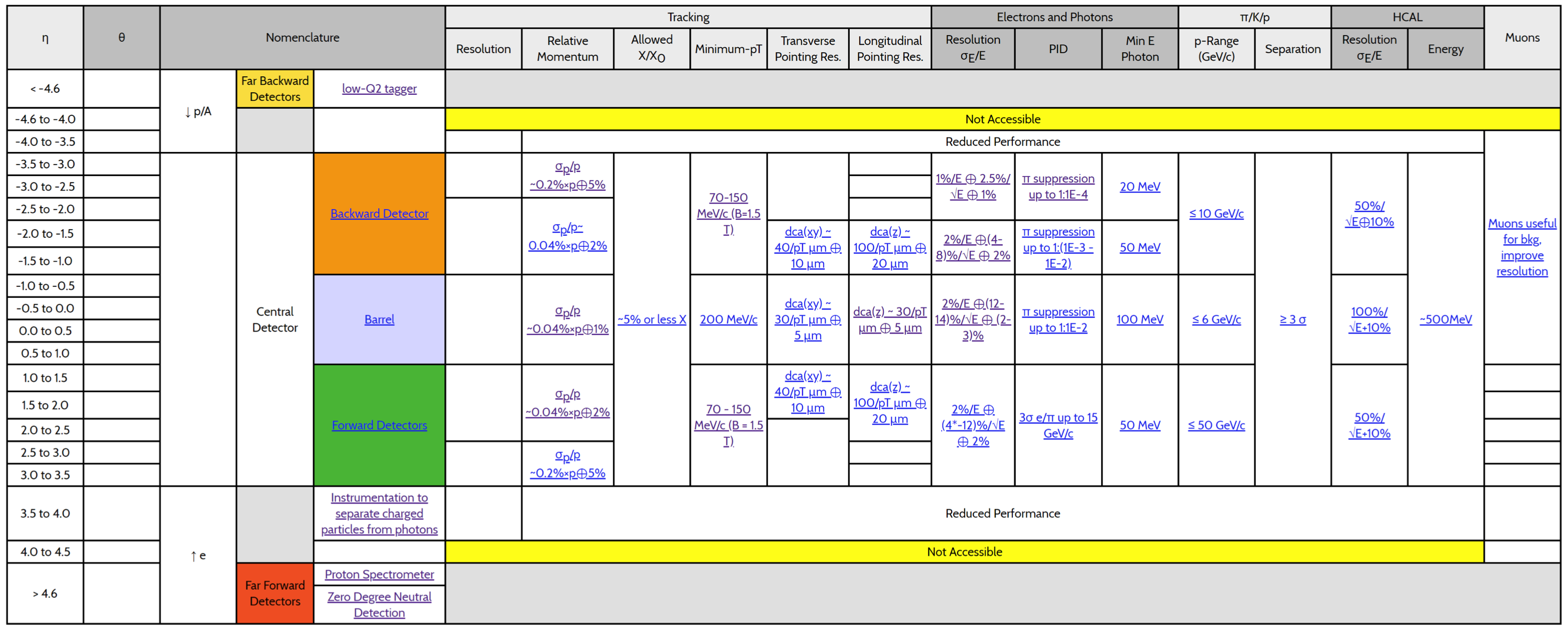}
\caption{\label{fig:pid_backbone} (left) The imaging Cherenkov detectors for particle identification in a proposed EIC detector concept;%. Image taken from \cite{fanelli_ai4cherenkov}; 
 (right) the table with the coverage in momentum for particle identification taken from the EIC Yellow Report \cite{khalek2021science}.
}
\end{figure}

 All  detector designs proposed for EIC have a dual radiator ring-imaging Cherenkov detector (dRICH) in the hadron direction, detection of internally reflected Cherenkov light (DIRC) in the barrel, and a modular-aerogel RICH (mRICH) in the electron direction (as displayed in Fig.~\ref{fig:pid_backbone}).\footnote{Cf. presentations by the proto-collaborations ATHENA, CORE and ECCE at the EICUG Summer 2021 \cite{eicug_summer_2021}.}  
 
As discussed in \cite{joosten_simulations}, the simulation of Cherenkov detectors involve optical processes with many photons that need to be tracked through complex surfaces, making these detectors relatively slow  to simulate (CPU intensive) with full simulations like Geant \cite{ agostinelli2003geant4}:
(i) the mRICH, for example, has photons that originate in the aerogel and pass through a Fresnel lens made by many grooves whose impact on the simulation performance is non-negligible (for more details, see, \textit{e.g.}, \cite{he2020development}); 
(ii) The dRICH uses an aerogel radiator and a large volume heavy gas radiator: light needs to propagate to mirrors and eventually to light sensors, and nested ring patterns with noise are utilized for particle identification (more details can be found in \cite{cisbani2020ai});
(iii) The DIRC uses long quartz bars coupled to an expansion volume: it has similar challenges with a much more complex optics system resulting in more complex hit patterns (more details can be found in \cite{kalicy2018high}).

\begin{figure}[!htbp]
\centering 
\includegraphics[width=.80\textwidth]{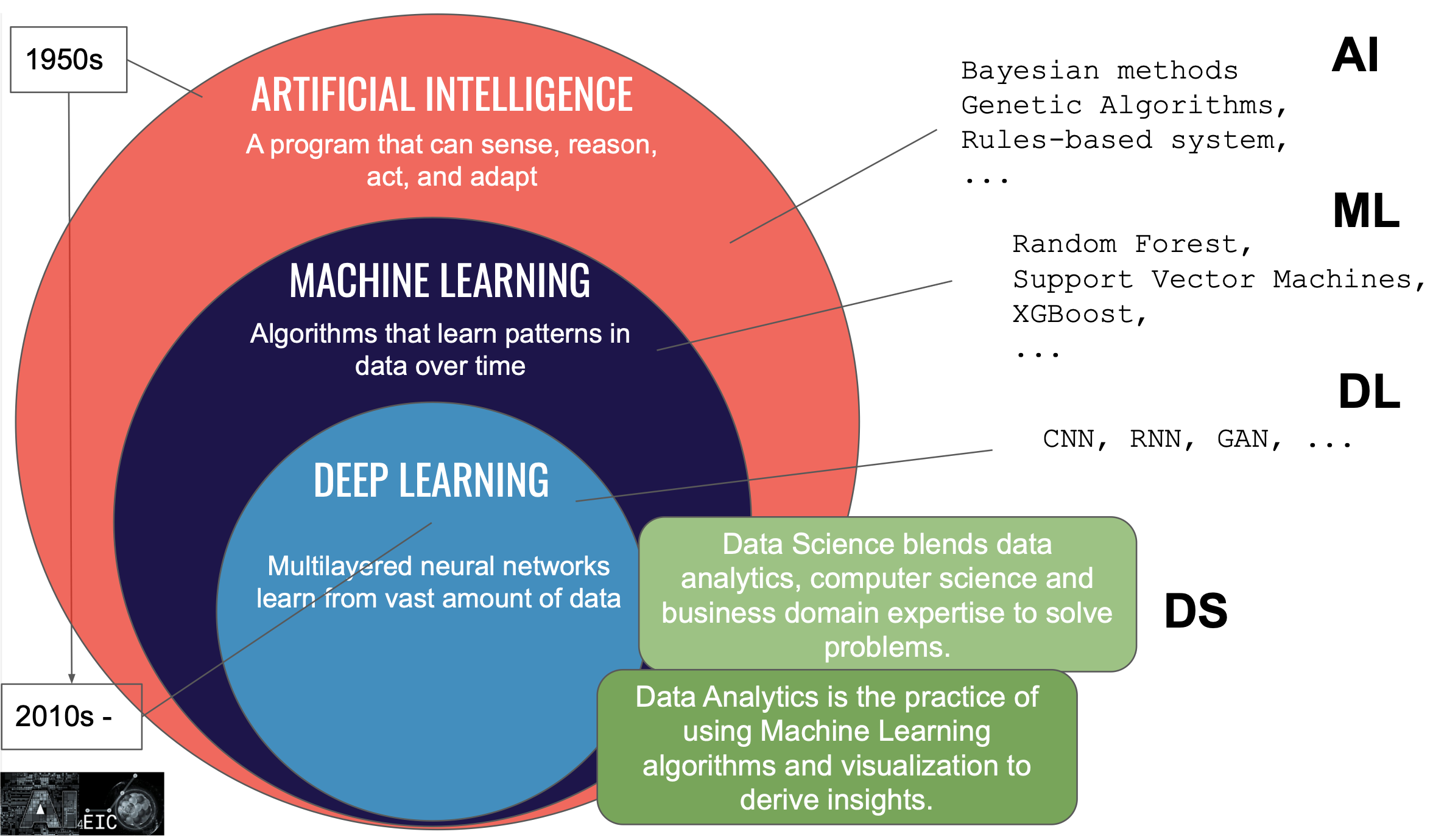}
\caption{\label{fig:AI} Taxonomy followed at AI4EIC defining AI, and ML and DL as subsets of AI. %Image taken from \cite{fanelli_ai4eic_structure}.
}
\end{figure}

Following the taxonomy of Fig.~\ref{fig:AI}, Artificial Intelligence is already contributing to face challenges associated to computationally intensive simulations and complex pattern recognition problems, and in what follows we discuss how AI can play a role for imaging Cherenkov detectors that will be built at the Electron Ion Collider.

In Sec. \ref{sec:ai} we will describe recent activities involving AI applications for imaging Cherenkov detectors that can be utilized at the EIC; in Sec.~\ref{sec:conclusions} we present our conclusions and perspectives.

\section{AI for Imaging Cherenkov Detectors: Recent Activities}\label{sec:ai}

An ongoing effort in the EIC community is providing a detector concept that meets the physics requirements described in the EIC Yellow Report \cite{khalek2021science}.
The proposed design may be further optimized after the detector proposal submission. 
In Sec.~\ref{sec:dRICH_optimization} we will describe the opportunity of further optimizing the complex design of the dual RICH detector with AI.  
As we mentioned, imaging Cherenkov detectors are characterized by computationally
intensive simulations as well as complex patterns for particle identification. The most complex ring topologies are those of the DIRC detector, and in Sec.~\ref{sec:DIRC_deeplearning} we will describe recent works based on deep learning for the DIRC, which in principle can be also extended to other imaging Cherenkov detectors. %%

\subsection{Detector design assisted by AI: the dRICH case}\label{sec:dRICH_optimization}

Detector optimization is an essential part of the R\&D and design process that involves mechanical design and budget to realize the best performance possible.  
This process is anticipated to continue in the months following the detector proposal towards CD-2 and CD-3.

\begin{figure}[!]
    \centering
    \includegraphics[scale = 0.16]{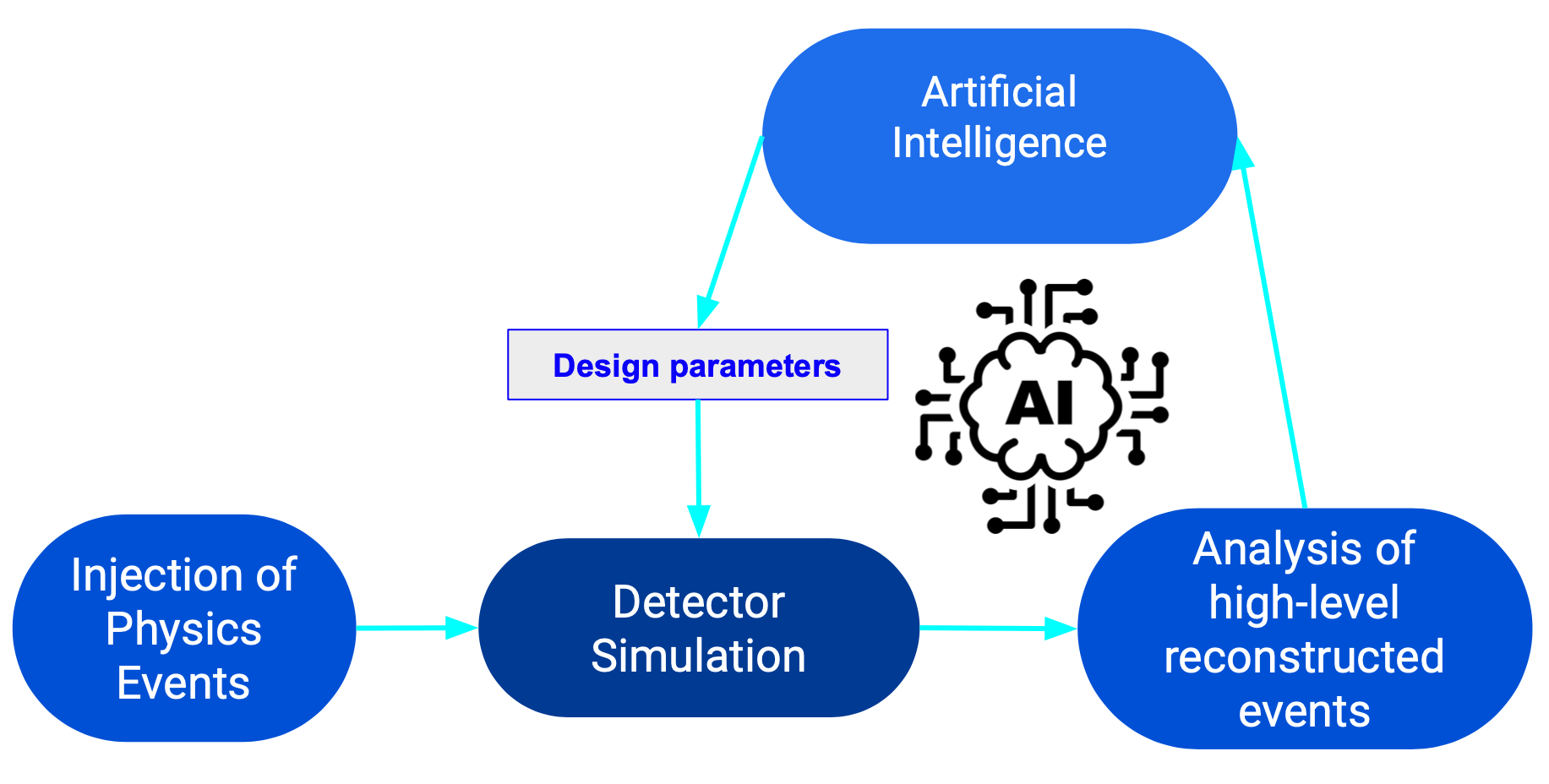}
    \caption{%(top) 
    Typical workflow of detector design assisted by AI: physics events are injected in a detector characterized by some given design parameters. Reconstructed events are analyzed and figures of merit are quantified and passed to some AI-based strategy, which in turn suggests the next design point to observe in this sequential approach; AI can also intervene in the simulation and reconstruction steps. 
    }
    \label{fig:design_AI}
\end{figure}

In general, a sequential AI-based strategy gathers the information associated to the proposed design point, \textit{i.e.} some figures of merit that quantify the goodness of the design, and based on this information suggests which design parameters to query at the next iteration (cf. workflow represented in Fig.~\ref{fig:design_AI}). 

This becomes particularly useful when facing with computationally intensive simulations, complex designs characterized by large dimensionality, and noisy black-box objective functions. 

The first parallelized, automated and self-consistent procedure for AI-optimized detector design was developed for the dRICH design at EIC by \cite{cisbani2020ai} leveraging Bayesian optimization (BO) \cite{snoek2012practical}: 
the baseline design consisted of two radiators (aerogel and C$_2$F$_6$ gas) sharing the same outward-focusing spherical mirror and highly segmented photosensors ($\approx 3$\,mm$^2$ pixel size) located outside of the charged-particle acceptance. The work in \cite{cisbani2020ai} was initially developed for the JLEIC design \cite{jleic} before Brookhaven National Laboratory (BNL) was selected for building the EIC, but the same exercise can be repeated for the dRICH design of detector concepts proposed by the proto-collaborations.

\begin{figure}[!ht]
\centering
\includegraphics[scale=0.30, angle = 0]{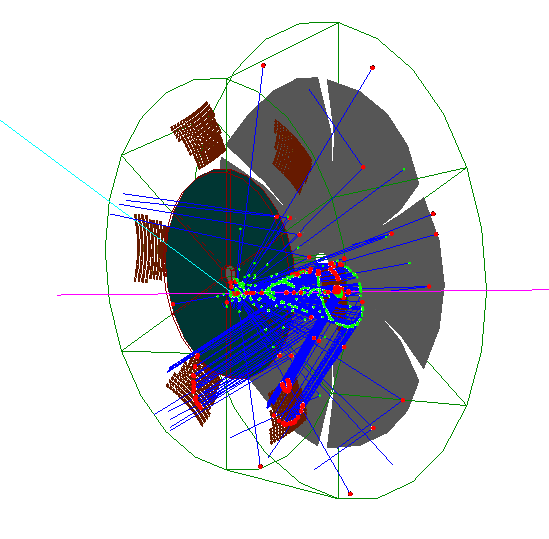}
\includegraphics[scale=0.32, angle = 0]{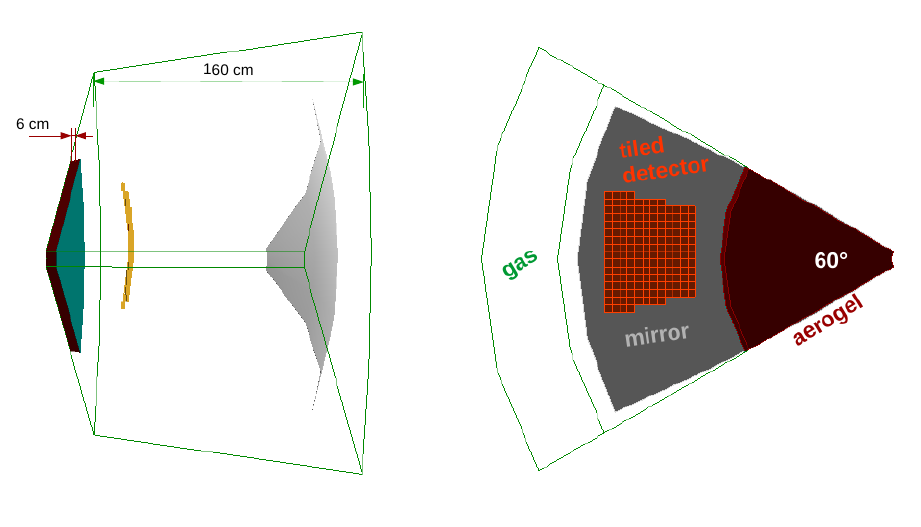}
\caption{Geant4 based simulation of the dRICH  \cite{cisbani2020ai} (left: entire detector; right: one of the six sectors). In transparent wired red is the aerogel radiator, in transparent wired green is the gas radiator volume; the mirrors sectors are in gray and the photo-detector surfaces (spherical shape) of about 8500~cm$^2$ per sector in dark-yellow. A pion of momentum 10~GeV/c is simulated \cite{cisbani2020ai}.
\label{fig:dual1}
}
\end{figure}

\begin{figure}[!]
\centering 
\includegraphics[width=.5\textwidth,origin=c,angle=0]{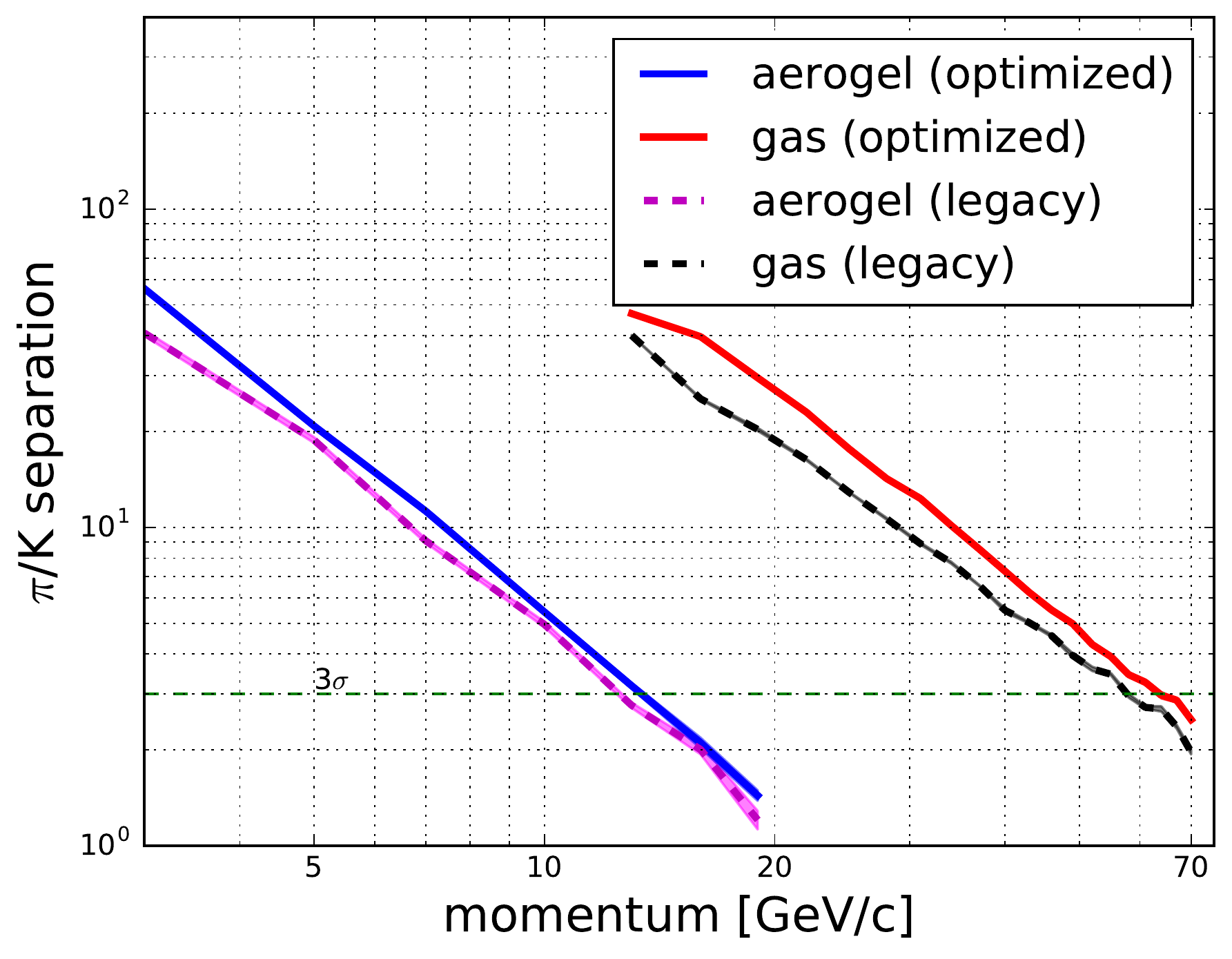}
\caption{\label{fig:drich_optimal} 
$\pi/K$ separation as number of $\sigma$, as a function of the charged particle momentum. 
The plot shows the improvement in the separation power with the approach discussed in \cite{cisbani2020ai} compared to the legacy baseline design \cite{del2017design}. The curves are drawn with 68\% C.L. bands.  
}
\end{figure}

The dRICH detector in the hadron endcap is essential for particle identification in a wide range of momentum, cf. table in Fig.~\ref{fig:pid_backbone} (right).  
In \cite{cisbani2020ai}, the important role played by certain parameters characterizing the design of the dRICH has been shown, particularly the mirror radius and its position, the location of the detecting tiles in each of the six modular petals of the dRICH, and the aerogel refractive index and thickness.
Results of the AI-based optimization are shown in Fig.~\ref{fig:drich_optimal} which shows the improvement in the $\pi$/K separation power.
Similarly the dRICH parameters can be fine-tuned in the designs proposed by the proto-collaborations ATHENA, CORE and ECCE, in each case considering the differences and constraints imposed by the global detector design. 

Noticeably, EIC is already utilizing AI-supported optimization of the detector for the ongoing detector proposal. ECCE, for example, has built a multi-objective optimization for the design of the tracking system, and more details can be found here \cite{phelps_ecce}. %fanelli_aimldesign, fanelli2021ai

\subsection{Deep learning for fast simulations and particle identification: the DIRC example}\label{sec:DIRC_deeplearning}

As already mentioned, simulation of imaging Cherenkov detectors involve optical processes with many photons that need to be tracked through complex surfaces, making simulations computationally intensive. 
In addition to that, detectors like the DIRC present complex hit patterns (for topology and sparsity of the hits) which can make difficult the extraction of information about the particle to identify. 
As discussed during the AI4EIC workshop, these features seem to offer a natural place for deep learning applications for fast simulations and reconstruction.

\begin{figure}[!]
\centering 
\includegraphics[width=.75\textwidth,origin=c,angle=0]{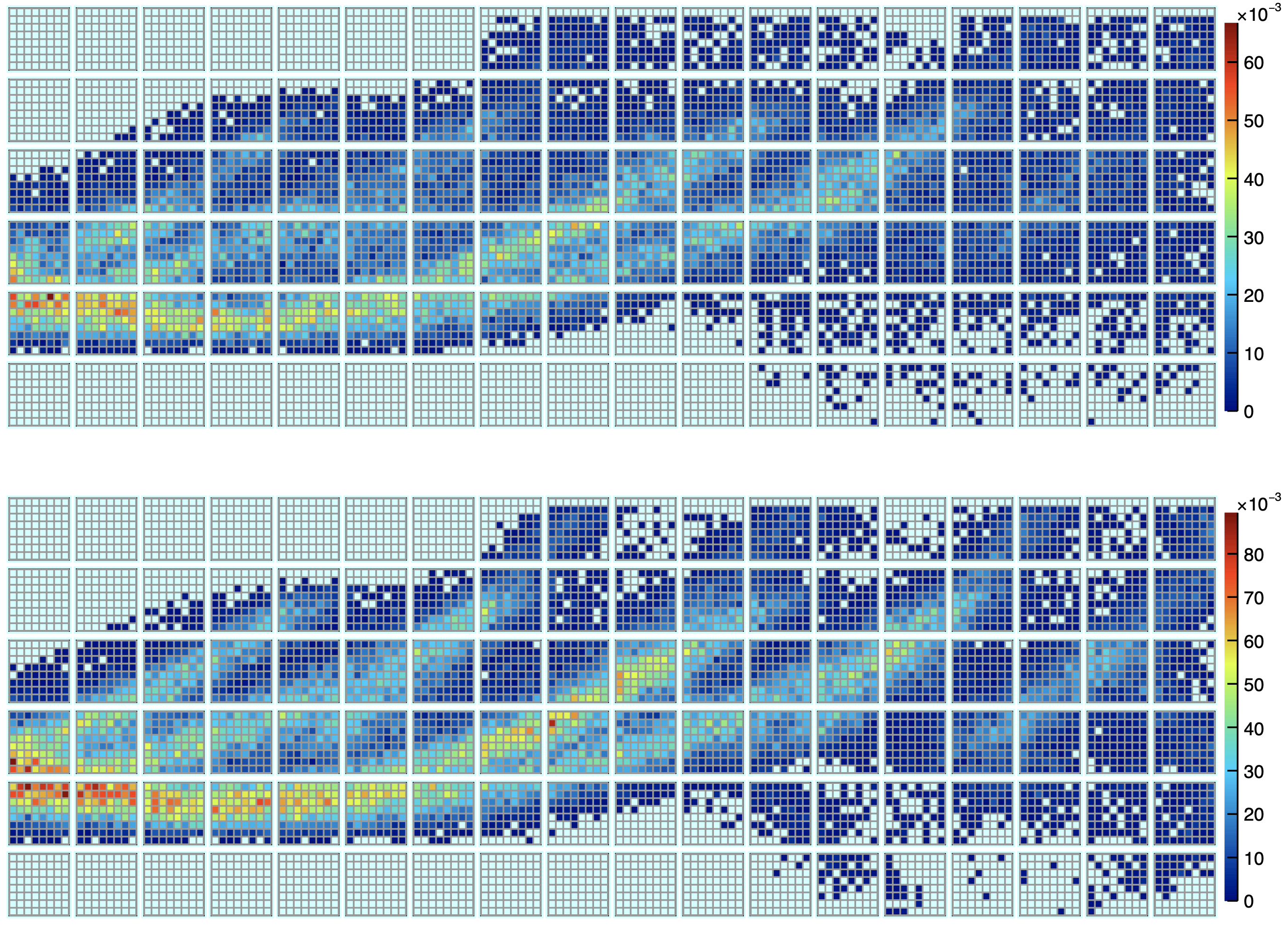}
\caption{\label{fig:DIRC_EIC} 
Example of complexity of hit patterns for the \GlueX \ DIRC detector \cite{ali2020gluex}. Hit pattern for $\pi$+ track: real data (top) and GEANT MC simulation (bottom).
}
\end{figure}

In what follows we will focus on the DIRC detector, for which there are already developed examples and ongoing activities using AI. 
%
%The different designs of the DIRC for the three proto-collaborations are reported in Fig.~\ref{fig:DIRC_EIC}, which also displays a hit pattern obtained with charged tracks of pions at 6 GeV/c in momentum.  
%
Fig.~\ref{fig:DIRC_EIC} displays an example of a hit pattern obtained with charged tracks of pions detected by the \GlueX \ DIRC detector \cite{barbosa2017gluex}.  

Generative Adversarial Networks (GANs) \cite{goodfellow2014generative} have been used to simulate the response of the \GlueX \ DIRC in a work by \cite{derkach2019cherenkov} to bypass low-level details at the photon generation stage. 
The architecture is trained to reproduce high-level features of the incident charged particles simulated with FastDIRC \cite{hardin2016fastdirc}, and allows for a dramatic increase of simulation speed.\footnote{The FastDIRC package \cite{hardin2016fastdirc} allows for fast simulation of the hit patterns as well as PID through a likelihood-based approach.}

GANs have been also recently used in LHCb for event generation and simulation of detector responses \cite{anderlini_LHCb_simulations}. 
In fact, the increasing luminosities of future LHC runs will require an unprecedented amount of simulated events to be produced. The accurate simulation of Cherenkov detectors takes a sizeable fraction of CPU time and as an alternative high-level reconstructed observables can be generated with GANs to bypass low level details. In \cite{maevskiy2020fast}, in particular, the fast simulation is trained using real data samples collected by LHCb during run 2. 

A novel architecture for Deeply learning the Reconstruction of Imaging CHerenkov (DeepRICH) detectors directly from low level features has been proposed in \cite{Fanelli_2020}. 

\begin{figure}[!t]
  \centering
  \begin{tabular}[b]{cc}
    \begin{tabular}[b]{c}
      \begin{subfigure}[b]{0.3\columnwidth}
        \includegraphics[width=\textwidth]{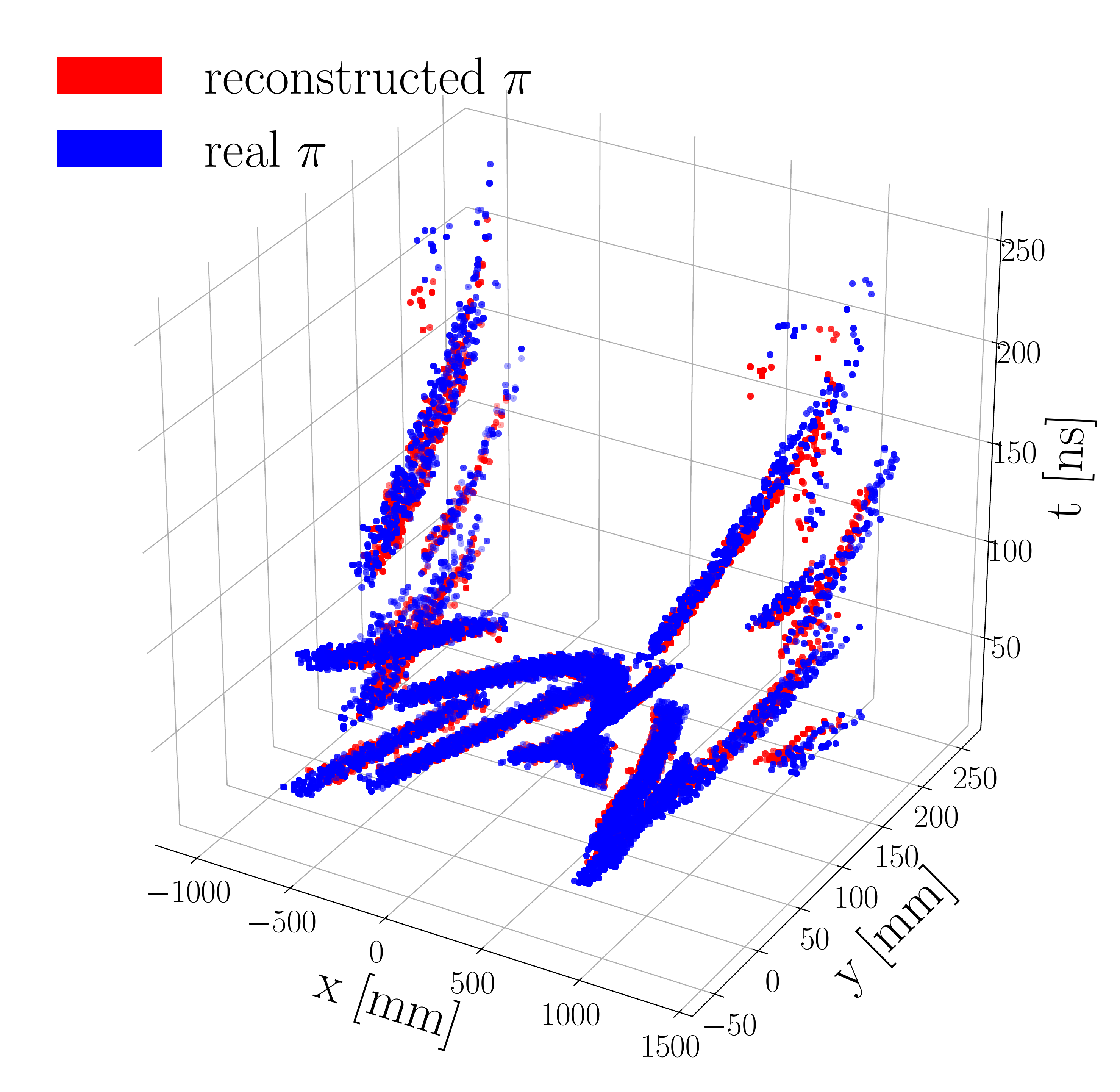}
        \caption{Example of hit points reconstructed by DeepRICH at 4 GeV/c, with an almost perfect overlap between the reconstructed and the injected hits of both pions. Image taken from \cite{fanelli2020machine}.}
        \label{fig:A}
      \end{subfigure}\\
      \begin{subfigure}[b]{0.3\columnwidth}
        \includegraphics[width=\textwidth]{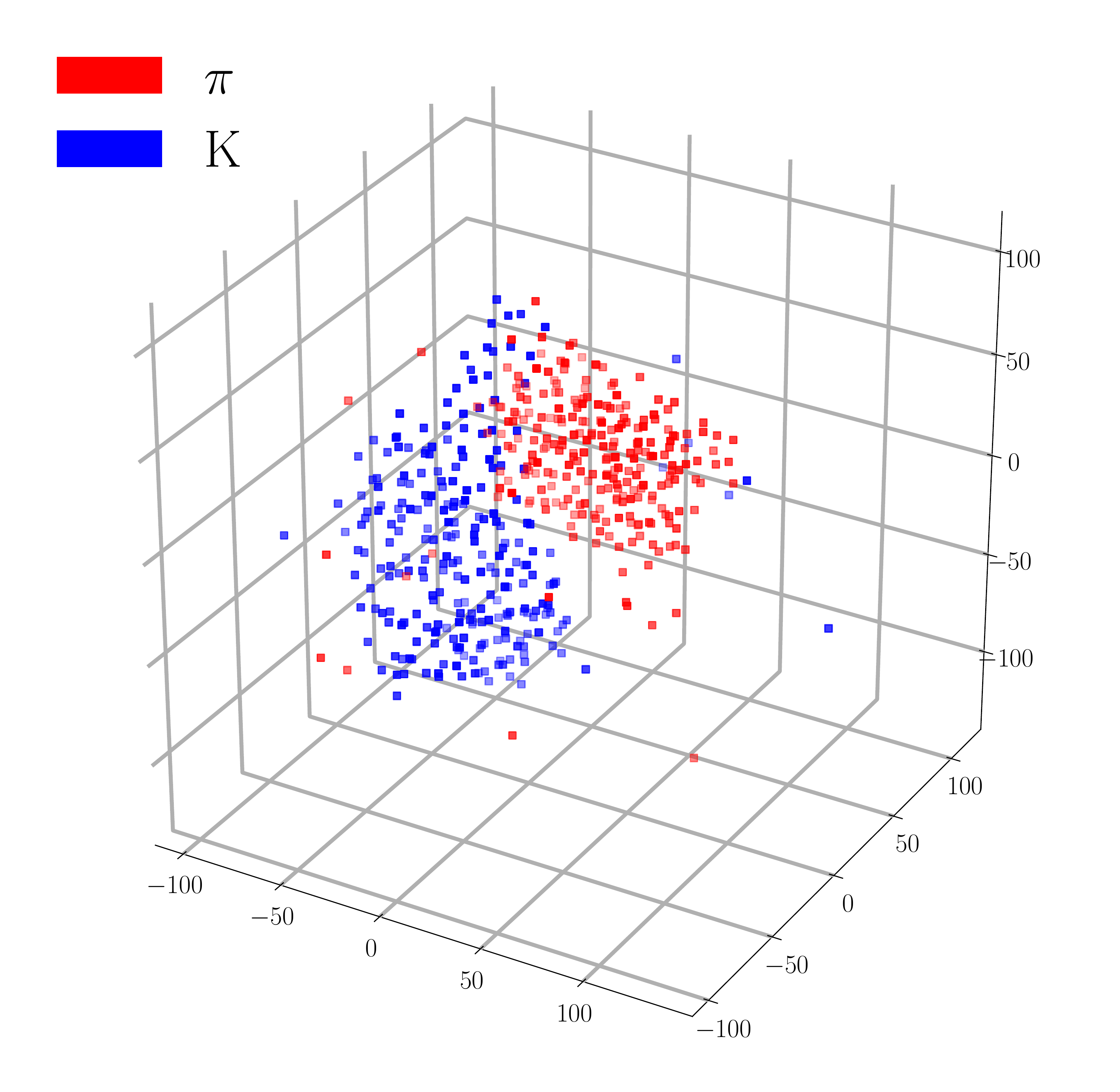}
        \caption{Example of distributions in the latent space for pions and kaons at 4 GeV/c. Image taken from \cite{fanelli2020machine}.}
        \label{fig:B}
      \end{subfigure}
    \end{tabular}
    &
    \begin{subfigure}[b]{0.45\columnwidth}
      \includegraphics[width=\textwidth]{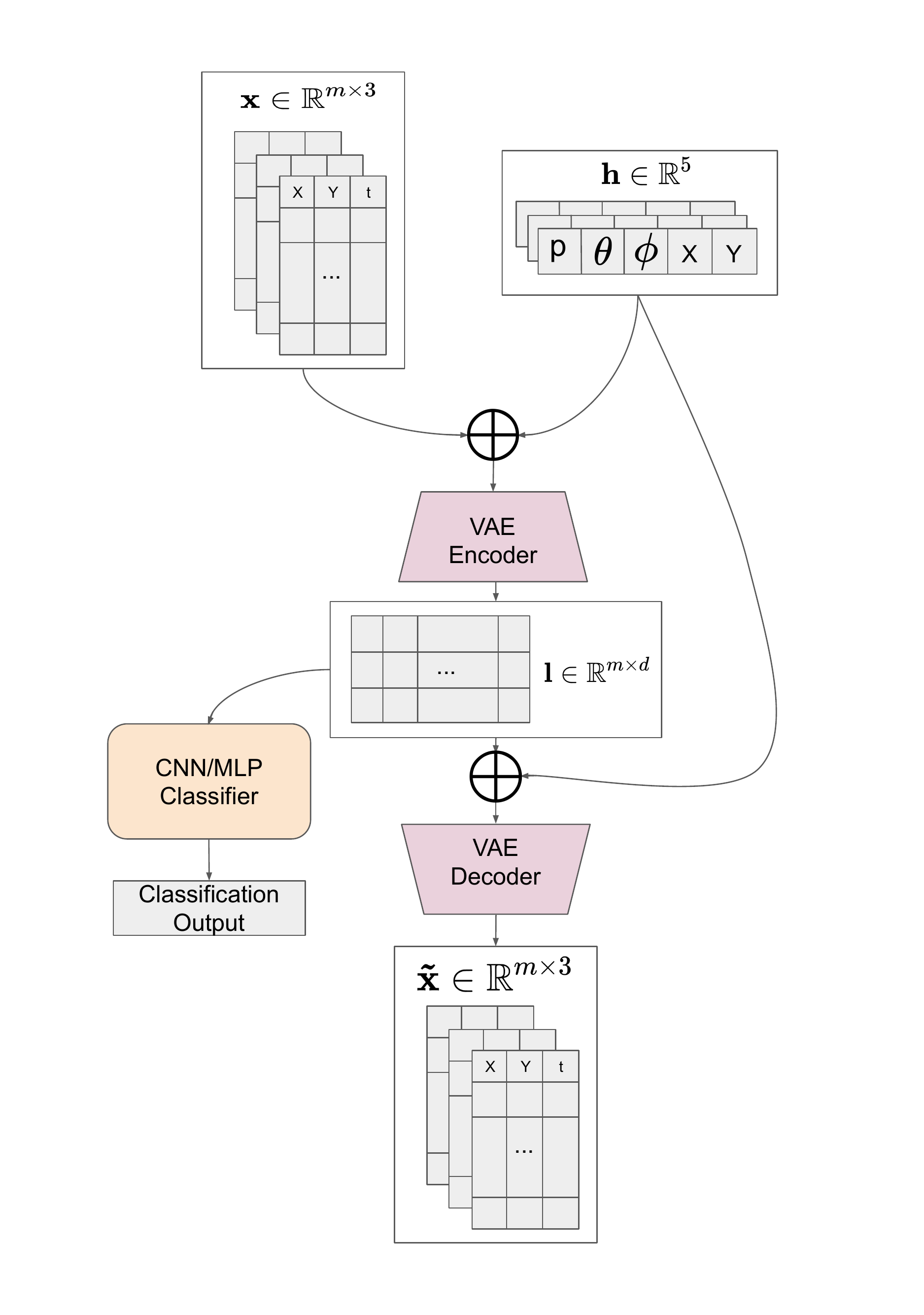}
      \caption{A flowchart of DeepRICH: the inputs are concatenated with the kinematics. VAE  generates a set of vectors of latent variables, which are then used for both the classification of the particle and for the reconstruction of the hits. Image taken from \cite{Fanelli_2020}.}
      \label{fig:C}
    \end{subfigure}
  \end{tabular}
  \caption{The DeepRICH architecture \cite{Fanelli_2020}. (a) an example of reconstructed hit patterns by the VAE; (b) an example of $\pi/K$ distinguishing power in the latent space at 4 GeV/c; (c) the DeepRICH flowchart: the CNN/MLP is used for particle identification based on the latent space distributions from the VAE.
  \label{fig:DeepRICH}}
\end{figure}

 A flowchart of DeepRICH is shown in Fig.~\ref{fig:DeepRICH}: it is a custom architecture consisting of Variational Autoencoders (VAE) \cite{kingma2019introduction} for reconstruction, and Convolutional Neural Networks (CNN) \cite{koushik2016understanding} combined with a Multilayer Perceptron (MLP) for particle identification.

The classification is supervised and needs labeled data. In \cite{Fanelli_2020} studies have been performed utilizing samples produced with FastDIRC for the \GlueX \ DIRC design. 
In \cite{fanelli_ai4cherenkov} it has been discussed: (a) the possibility of using for training high purity samples directly from real data using specific topologies with, \textit{e.g.}, $\pi$, K and p; (b) a potential procedure for data augmentation at any given bin of the particle kinematics, consisting of sampling the expected hit pattern according to the expected photon yield distribution.\footnote{This works if the hits forming the hit pattern can be considered independent from each other in good approximation .}  

 The main features of DeepRICH can be summarized in the following points: (i) it is fast and provides accurate reconstruction; 
 (ii) it can be extended to multiple particle types (multi-class identification); (iii) it can be generalized to fast simulation, using VAE as a generative model; (iv) it can utilize (x,y,t) patterns if time is measured; can deal with different topologies and different detectors; (v) it deeply learns the detector response (high-purity samples of real data can be injected during the training process). 

In terms of performance, it has been shown that the reconstruction efficiency is consistent with that of established methods such as FastDIRC \cite{hardin2016fastdirc}, with an area under the curve (AUC) close to that of FastDIRC across the entire phase-space that has been studied, namely AUC(DeepRICH) $\gtrsim$ 0.99 AUC (FastDIRC).
The main advantage is in the effective inference time per particle, which on a Titan V GPU resulted to be $\lesssim$ 1 $\mu$s utilizing a batch of 10$^{4}$ particles.  

DeepRICH has been prototyped looking at pions and kaons at \GlueX \ in a limited phase-space between 4 and 5 GeV and utilizing one fused silica bar. 
Exciting work is planned to extend the kinematics of the particle and the number of bars, as well as to improve the training time with the possibility of distributed training. 
Another interesting activity is that of utilizing architectures like DeepRICH as generative models for simulations of the hit patterns as a function of the kinematics. 
 Another potential application could be training DeepRICH using pure samples of identified particles from real data, allowing to deeply learn the response of the Cherenkov detector \cite{Fanelli_2020}.

\section{Conclusions and Perspectives}\label{sec:conclusions}

The last few years have been characterized by a groundswell of applications in nuclear and particle physics based on AI/ML both for fast simulations and for particle reconstruction and identification. 
The particle identification at the Electron Ion Collider is based on imaging Cherenkov detectors, which typically entail computationally intensive simulations and challenging pattern recognition problems.   
During the design phase, Artificial Intelligence can be utilized for optimizing the design of Cherenkov detectors and for reducing the computing budget necessary to explore a large number of design points. 
We also discussed how simulation speed up can be obtained using generative models like GANs, and covered novel multi-purpose architectures which can in principle be utilized for fast simulation, reconstruction and identification of particles. 
Architectures like DeepRICH can work for different imaging Cherenkov detectors and topologies of hit patterns, and lots of exciting activities are planned in the next few years to extend and characterize the performance of these applications.

\bibliography{references}
\end{document}